\documentclass[]{beilstein}
\usepackage{xspace}
\usepackage{hyperref}
\usepackage{xcolor}
\hypersetup{
    colorlinks,
    linkcolor={red!50!black},
    citecolor={blue!50!black},
    urlcolor={blue!80!black}
}
\usepackage[separate-uncertainty=true]{siunitx}
\newcommand{\meff}{\ensuremath{ m_{\mathrm{eff}} }}
\newcommand{\Fmin}{\ensuremath{ F_{\mathrm{min}} }}
\newcommand{\kB}{\ensuremath{ k_{\mathrm{B}} }}
\newcommand{\kex}{\ensuremath{ \kappa_\mathrm{ex} }}
\newcommand{\Svv}{\ensuremath{S_\mathrm{VV} }}
\newcommand{\Szz}{\ensuremath{ S_\mathrm{\zeta\zeta} }}
\newcommand{\nc}{\ensuremath{ n_\mathrm{c} }}
\newcommand{\Wm}{\ensuremath{ \Omega_\mathrm{m} }}

\begin{document}
\nolinenumbers

\title{Low-temperature AFM with a microwave cavity optomechanical transducer}
\author*[1]{Ermes Scarano}{ermes@kth.se}
\author[1]{Elisabet K. Arvidsson}
\author[1]{August K. Roos}
\author{Erik Holmgren}
\affiliation{Department of Applied Physics, KTH Royal Institute of Technology, Hannes Alfvéns väg 12, SE-114 19 Stockholm, Sweden}

\author[2]{Riccardo Borgani}
\author{Mats O. Tholén}
\affiliation{Intermodulation Products AB, SE-823 93 Segersta, Sweden }

\author[1]{David B. Haviland}

\maketitle

\begin{abstract}
We demonstrate AFM imaging with a microcantilever force transducer where an integrated superconducting microwave resonant circuit detects cantilever deflection using the principles of cavity optomechanics. 
We discuss the detector responsivity and added noise pointing to its crucial role in the context of force sensitivity. 
Through analysis of noise measurements we determine the effective temperature of the cantilever eigenmode and we determine the region of detector operation in which the sensor is thermal-noise limited.   
Our analysis shows that the force-sensor design is a significant improvement over piezoelectric force sensors commonly used in low-temperature AFM.
We discuss the potential for further improvement of the sensor design to achieve optimal detection at the standard quantum limit.  
We demonstrate AFM operation with surface-tracking feedback in both amplitude-modulation and frequency-modulation modes.  
\end{abstract}


\section{Introduction}

The past two decades have seen the emergence of a variety of remarkable microscopic and mesoscopic optomechanical devices. 
Through innovative design and fabrication, new record values of fundamental figures of merit have been reported at a surprisingly rapid rate. 
The field is now mature for the next challenge -- application-ready devices. Optomechanical interaction has been proposed to enable or improve many applications~\cite{metcalfe2014applications}, including
accelerometers~\cite{krause2012high},
tests of quantum gravity~\cite{pikovski2012probing,marletto2017gravitationally,krisnanda2017revealing}, 
force microscopy~\cite{halg2021membrane,liu2012wide}, magnetometry~\cite{forstner2014ultrasensitive} and quantum state transfer~\cite{hill2012coherent}.
In some cases further improvement on fundamental figures of merit is required, while in other cases the difficulty lies in balancing trade-offs to find an optimal design which fulfills the specific requirements of the application. 
The latter is indeed the case for force sensing in atomic force microscopy (AFM).
Force transduction at maximum sensitivity requires detecting the position of a ``test mass'', while minimizing the added noise introduced by the detection itself~\cite{braginskiui1975quantum,smith1995limits}.
The challenge for high-resolution AFM is designing such a detector for a force transducer hosting a sharp tip and scanning over a surface in ultra-high vacuum and at ultra-low temperature.

In this paper we report on an AFM cantilever force sensor with an integrated superconducting microwave resonant circuit.
Using the principles of cavity optomechanics~\cite{aspelmeyer2014cavity}, we detect the deflection of the cantilever from its equilibrium position as a shift of the microwave resonance.
We briefly summarize the operating principle of the force sensor based on kinetic-inductive electromechanical coupling (KIMEC), whose design and fabrication were already presented in detail in previous publications by our group \cite{roos2023kinetic,roos2024design}.
Here we focus on the deployment of the force sensor, demonstrating force-gradient sensing and scanning over a test surface at \SI{10}{\milli\kelvin} in a closed-cycle dilution refrigerator (DR). 
We operate the microscope with surface-tracking feedback using the two most common imaging modes of dynamic AFM: amplitude modulation (AM-AFM) and frequency modulation (FM-AFM). 

One of the biggest challenges of operating an AFM in a closed-cycle DR is the pulse tube cryogenic head causing significant wideband mechanical vibrations which can propagate through the cold stages down to the measurement apparatus.
However, force sensors with integrated position detectors, as the one presented in this work, do not require free-space optical alignment with a long optical path length. 
They can therefore be made light and compact, significantly reducing their susceptibility to external vibrations, the complexity of vibration isolation, and their thermal mass.
For this prototype we implement a rather simple vibration isolation which, as we will show, is sufficient to demonstrate imaging features in a \SI{1}{\micro\meter} scan field, but has not been tested to atomic resolution.

\section{Force sensitivity and imaging}

The microcantilever in dynamic AFM operates as a resonant force transducer. Each individual eigenmode is described by a mechanical susceptibility $\chi$, expressing cantilever deflection $\zeta$ in response to a force $F$ at the frequency $\Omega$
\begin{equation}
    \chi(\Omega)=\frac{\zeta(\Omega)}{F(\Omega)}=\frac{1}{ \meff ( \Wm^2-\Omega^2+i\Omega\Gamma)},
    \label{susceptibility}
\end{equation}
where $\Wm=\sqrt{k/\meff}$ is the resonance frequency and $\Gamma = \eta / \meff$ the damping rate, with \meff, $k$ and $\eta$ respectively the effective mass, stiffness and damping coefficient of the eigenmode.

A key figure of merit of a force transducer is its force sensitivity determined by all sources of noise.
The total noise can be expressed as the sum of the power spectral densities (PSD) of uncorrelated contributions such as the thermal fluctuations in cantilever deflection $\Szz^\mathrm{cant}(\Omega)$ [\unit[per-mode=symbol]{\meter\squared\per\hertz}] at the effective mode temperature $T$, and the frequency-independent added noise of the detection \Svv [\unit[per-mode=symbol]{\volt\squared\per\hertz}] which is converted to an equivalent deflection noise $\Szz^\mathrm{det}$ via the detector responsivity $\alpha$ [\unit[per-mode=symbol]{\volt\per\meter}]
\begin{equation}
    \Szz^\mathrm{tot}(\Omega)=\Szz^\mathrm{det}+\Szz^\mathrm{cant}(\Omega)=\frac{\Svv}{\alpha^2}+2k_BT\eta|\chi(\Omega)|^2.\label{Stot}
\end{equation}
The noise-equivalent force for a given measurement bandwidth $\Delta f$ is given by,
\begin{equation}
    F_\mathrm{N}(\Omega)=\left(~\int^{\Omega+\pi\Delta f}_{\Omega-\pi\Delta f}\frac{\Szz^\mathrm{tot}(\Omega)}{|\chi(\Omega)|^2}\frac{d\Omega}{2\pi}\right)^{1/2}.
\end{equation}
This quantity is minimized at the mechanical resonance frequency, where $|\chi(\Omega_m)|^2=\meff / k\eta^2$, resulting in a minimum detectable force,
\begin{equation}
    \Fmin=F_\mathrm{N}(\Wm)\approx \sqrt{\left(\frac{\Svv k\eta^2}{\alpha^2 \meff }+2k_BT\eta\right)\Delta f}. \label{F_min}
\end{equation} 

It is worth noting that \cref{F_min} clarifies a recurring misconception. Decreasing stiffness $k$ does not necessarily reduce $F_\mathrm{min}$. Rather, it relaxes the constraint on the detector's added noise for the force sensor to operate in the thermal-limited regime $\Szz^\mathrm{det}\ll\Szz^\mathrm{cant}(\Omega_m)$, where $F_\mathrm{min}\simeq \sqrt{2\kB T\eta\Delta f}$. Achieving this limit can be especially challenging at cryogenic temperatures.

Mamin \emph{et al.} achieved a record force sensitivity of $\SI{0.82}{\atto\newton\per\sqrt{\hertz}}$ using a cantilever with  $\Wm=\SI{5}{\kilo\hertz}$ and stiffness $k=\SI[per-mode=symbol]{260}{\micro\newton\per\meter}$ at $T=\SI{110}{\milli\kelvin}$~\cite{mamin2001sub}.
AFM usually requires stiffness on the order of \SI[per-mode=symbol]{100}{\newton\per\meter} to avoid the tip jumping to contact and sticking to the surface~\cite{garcia2002dynamic}.
Such stiffness would necessitate five orders of magnitude improvement in detector noise to achieve an equivalent force sensitivity. Reduction of detection noise is therefore critical for improving low temperature AFM.

\section{AFM sensor and detection principle}

\begin{figure}
\centering
\caption{(a)--(c) Scanning electron micrograph of the probe, featuring a Si-N triangular cantilever released from the Si substrate, hosting an integrated lumped-element LC resonator realized from a single layer of Nb-Ti-N and a Pt-C tip deposited at the cantilever apex. (d) Experimental setup. The scanner box is suspended from the \SI{10}{\milli\kelvin} stage with soft springs. Folded copper braids provide thermalization and additional mechanical damping. The digital synthesizer and analyzer (\emph{Presto} platform by Intermodulation Products AB) provides the DC output to control the scanner, the radio-frequency drive for the piezo shaker and the microwave multifrequency lock-in modulation/demodulation.}
\label{fig:probe}
\dblcolfigure{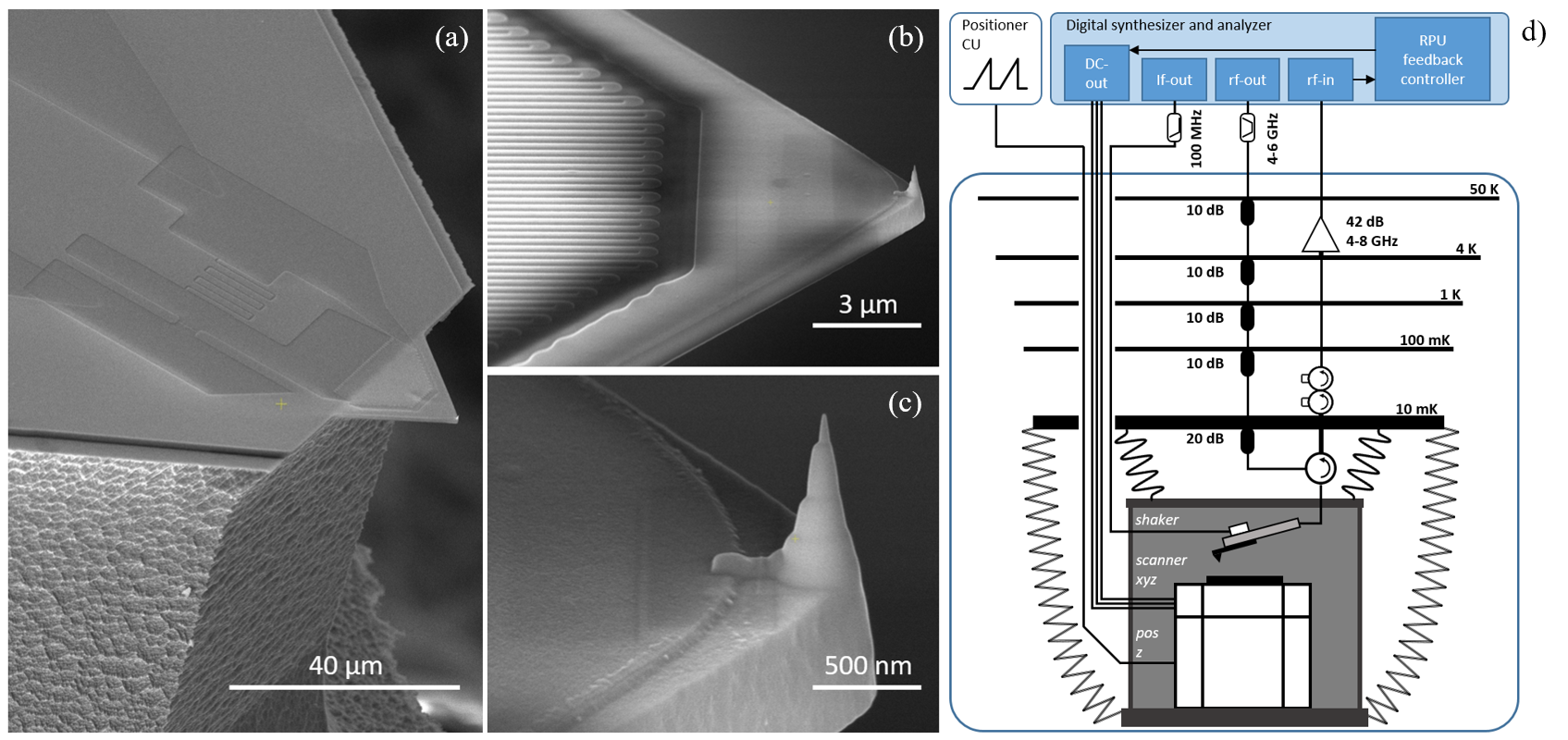}
\end{figure}

The advantages of cavity optomechanical detection for AFM sensors were already showcased by Liu \emph{et al.}~\cite{liu2012wide}, who implemented a whispering gallery mode toroidal optical cavity coupled through its evanescent field to a doubly-clamped beam with a tip in the middle. 
We replace the optical cavity with a superconducting microwave resonant circuit which is comparatively easy to fabricate and to integrate on an AFM cantilever with a tip at the free end.
\cref{fig:probe}(a)--(c) shows a scanning electron micrograph of the probe. 
The triangular cantilever is a Si-N plate released from a Si substrate.
The microwave superconducting lumped-element resonant circuit consists of an inter-digital capacitor in series with a meandering nanowire inductor, both fabricated from a single layer of Nb-Ti-N deposited on the Si-N. 
The nanowire meanders across the clamping edge of the plate, giving rise to a modulation of its kinetic inductance due to surface strain generated by cantilever bending.
A Pt-C tip with curvature radius $<\SI{10}{\nano\meter}$ at its apex is formed at the free end of the cantilever through a series of electron-beam-assisted depositions.

\cref{fig:probe}(d) shows a schematic of the experimental setup implementing a standard low-noise cryogenic microwave reflection measurement of the sensor, which is mounted in a prototype AFM suspended with three springs from the \SI{10}{\milli\kelvin} plate of the DR. 
The microscope consists of a metallic box which supports the scanner and a printed-circuit board to which the sensor is mounted. 
This custom board also hosts a piezoelectric inertial actuator (shaker) which drives the cantilever oscillation.
A test surface is mounted on top of the open-loop scanner stack consisting of a $z$-positioner for coarse approach and an open loop $xyz$-scanner for fine positioning (Attocube anpz 102/LT/HV, ansxyz 100/LT/HV). 
Twisted pairs and coaxial cables are coiled to make springy electrical connections to the AFM.  
Thermal anchoring of the various parts is achieved by copper braids with multiple zig-zag folds.  
At room temperature the resonance frequency of the AFM's suspension system is roughly \SI{1}{\hertz}, with a quality factor of roughly 2.

Mechanical oscillation of the tip causes phase modulation of the reflected microwave pump, detected as motional sidebands in the  signal spectrum.
Measuring the microwave response at a sideband, the detection responsivity $\alpha$ [\unit[per-mode=symbol]{\volt\per\meter}] can be expressed as 
\begin{equation}
    \alpha=\sqrt{Z_0G^2 \left( \frac{P_\mathrm{in}}{\hbar\omega_\mathrm{p}}\frac{\kex}{\kappa^2/4+\Delta^2} \right)  \frac{\kex}{(\kappa/2)^2+(\Delta+\Omega)^2} }
    =\sqrt{Z_0G^2 n_\mathrm{c} \frac{\kex}{(\kappa/2)^2+(\Delta+\Omega)^2} }
    \label{responsivity}
\end{equation}
where  $G=\partial\omega_\mathrm{0}/\partial \zeta$ [\unit[per-mode=symbol]{\hertz\per\meter}] is the electromechanical coupling coefficient, $\kappa_{ex}$ the external loss rate and $\kappa$ the total loss rate of the microwave resonance, and $Z_0$ the characteristic impedance of the transmission line. 
The term in the first parenthesis is the circulating power or energy stored in the cavity, expressed as the intracavity photon number $n_\mathrm{c}$.
We control $\nc$ through the drive power of the pump $P_\mathrm{in}$ and the detuning from the cavity resonance frequency $\Delta=\omega_\mathrm{p}-\omega_\mathrm{0}$. 
In principle, arbitrarily large responsivity $\alpha$ is achieved for arbitrary weak coupling $G$ by increasing $P_\mathrm{in}$. 
In practice, nonlinear effects emerge at large $\nc$, such as heating (nonlinear loss) and resonance-frequency shift (Kerr nonlinearity), which cause the linear response picture of cavity optomechanics to break down.

Detection noise in the cavity optomechanical scheme is determined by quantum fluctuations of $\nc$, or photon shot noise.
Using \cref{responsivity} for the responsivity $\alpha$ we can convert this detection noise to an equivalent mechanical displacement noise or imprecision noise, inversely proportional to $\nc$,
\begin{equation}
     S_\mathrm{\zeta\zeta}^\mathrm{imp}(\Omega)=\frac{\zeta_\mathrm{zpf}^2}{4g_\mathrm{0}^2\nc}\left(\frac{\kappa^2/4+(\Omega+\Delta)^2}{\kex}\right).\label{imp_noise}
\end{equation}
Here we introduce the single photon coupling rate $g_0 = G  \zeta_\mathrm{zpf}$, corresponding to the microwave resonance frequency shift associated with the quantum zero-point motion of the mechanical mode $\zeta_\mathrm{zpf}=\sqrt{\hbar/2\meff\Wm}$.
The cavity optomechanical measurement also introduces a fluctuating radiation pressure force on the test mass or back-action noise, proportional to $\nc$~\cite{bowen2015quantum,schliesser2009resolved}
\begin{equation}
     S_\mathrm{\zeta\zeta}^\mathrm{BA}(\Omega)=\frac{\hbar^2g_\mathrm{0}^2\nc}{\zeta_\mathrm{zpf}^2}\left(\frac{\kex}{\kappa^2/4+(\Omega+\Delta)^2}\right)\left|{\chi(\Omega)}\right|^2.     
\end{equation}

When the microwave pump frequency is detuned by $\Delta=-\Wm$, and the mechanical mode is measured on resonance $\Omega=\Wm$ where $|\chi(\Wm)|=2\zeta^2_\mathrm{zpf}/\hbar\Gamma$, detector noise is minimized when $\Szz^\mathrm{imp}=\Szz^\mathrm{BA}$, where
\begin{equation}
   \nc=  \nc^\mathrm{SQL}=\frac{\Gamma}{4g_0^2}\frac{\kappa^2/4}{\kex}.
\end{equation}
If the mechanical oscillator is cooled to a very low temperature, so that it is in its quantum ground state, a measurement at $n_c^\mathrm{SQL}$ gives $\Szz^\mathrm{SQL}(\Wm)=2\zeta_\mathrm{zpf}^2/\Gamma$, the so-called standard quantum limit (SQL).
In an actual experiment components after the cavity in the measurement chain introduce additional imprecision noise, expressed as an added number of noise photons $n_\mathrm{add}$
\begin{equation}
    \Szz^\mathrm{det}=\Szz^\mathrm{imp}(1+2n_\mathrm{add})+\Szz^\mathrm{BA}\label{Sdet}.
\end{equation}
The added noise in our setup [\cref{fig:probe}(d)] is determined by the cryogenic low-noise amplifier (LNA) mounted at the \SI{4}{\kelvin} stage.  
We use Planck spectroscopy to calibrate our LNA, where we measure $n_\mathrm{add}\approx14$ at \SI{4.5}{\giga\hertz}~\cite{jolin2023multipartite,lingua2024continuous}. The addition of a near quantum-limited amplifier (e.g. a Josephson parametric amplifier) as the first stage in the amplification chain,  where  $n^\mathrm{add}\approx 1/2$, is a commonly adopted solution to reduce detection noise~\cite{Teufel2011}.

\section{Sensor characterization}

\cref{tab:1} summarizes the device parameters for three different probes employed in this work, measured at \SI{10}{\milli\kelvin} in free space far from the surface.
\begin{table}
\caption{Device parameters $\Omega_m$ and $\Gamma$ are determined by fitting the measured mechanical resonance.  $\omega_0$, $\kappa$ and \kex are determined by fitting to the measured microwave resonance, while \meff and $G$ are determined from FEM simulations~\cite{roos2024design}.}
\label{tab:1}
\begin{dblcoltabularx}{|l|l|l|l|l|l|l|X|}\hline
Probe  & \Wm~(\unit{\mega\hertz}) & $\Gamma$~(\unit{\hertz}) & \meff~(\unit{\pico\gram}) & $\omega_\mathrm{0}$ (\unit{\giga\hertz}) & $\kappa$~(\unit{\mega\hertz}) & \kex~(\unit{\mega\hertz}) &G~(\unit[per-mode=symbol]{\kilo\hertz\per\nano\meter})\\\hline
1 & 5.669 & 15 & 47 & 4.27 & 3.38 & 3.04 & 24.92 \\\hline 
2 & 4.963 & 5 & 61 & 4.64 & 1.03 & 0.64 & 24.56 \\\hline 
3 & 5.881 & 45 & 44 & 4.40  & 4.54 & 4.04 & 28.98 \\\hline 
\end{dblcoltabularx}
\end{table}

\begin{figure}
\centering
\caption{(a) Detector output noise equivalent displacement PSD and individual contributions.
(b) Noise contributions evaluated at the mechanical resonance as a function of the average circulating photon number $\nc$. The detector noise (solid red line) accounts for the intracavity ground state fluctuations (dashed red line), back action noise (dash-dot red line) and the added noise of the first stage of amplification. The horizontal line marks the fluctuations of the cantilever at its effective mode temperature. The solid and dashed black curves describe the total noise of the sensor for various cantilever mode temperatures.}
\label{fig:sensitivity}
\dblcolfigure{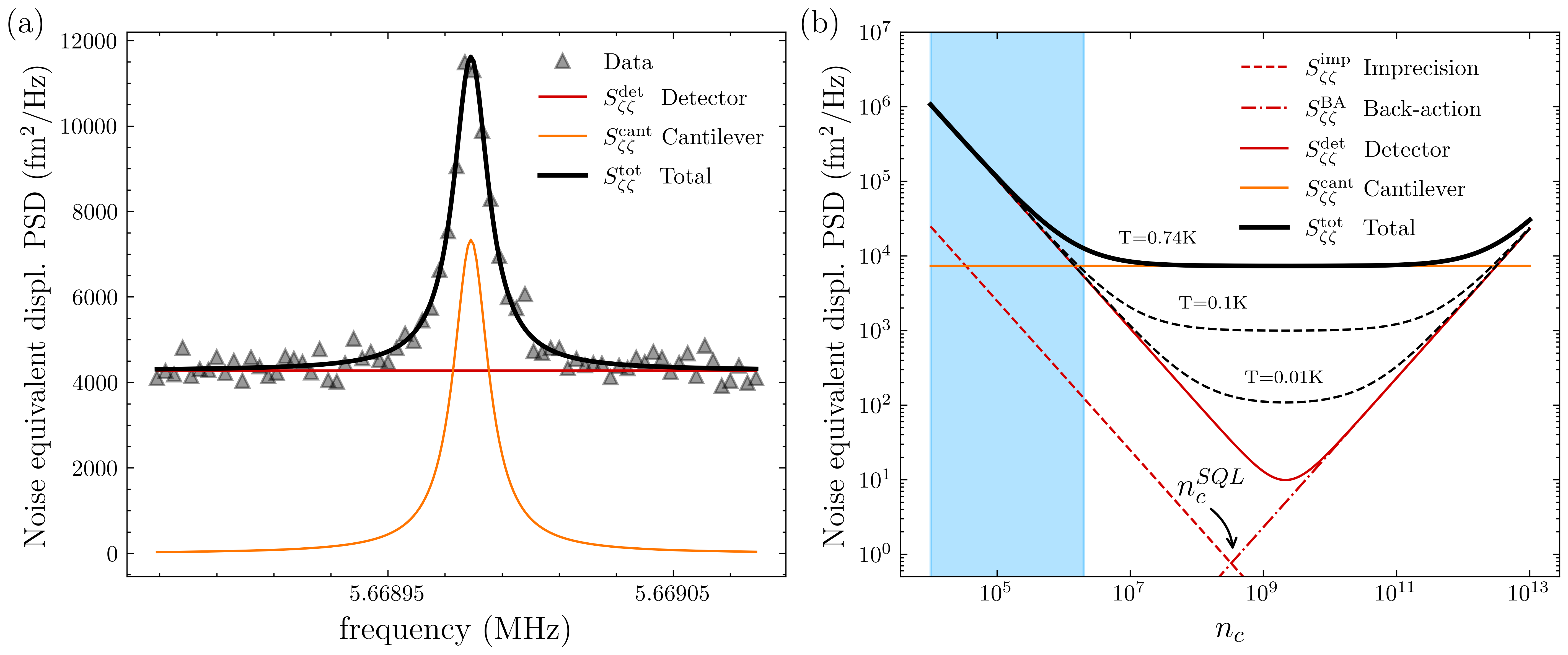}
\end{figure}

We characterize the sensor by measuring the total noise of the undriven mechanical resonance, up-converted to the microwave spectrum.
\cref{fig:sensitivity}(a) shows the data for Probe 1, obtained by converting the measured upper motional sideband to an equivalent displacement PSD $\Szz^\mathrm{tot}$ through \cref{responsivity} with $\Delta=-\Wm$ and $\nc=\num{1.98e6}$. 
Fitting \cref{Stot} to the measured data we extract the detector (red) and cantilever (orange) contributions to the total noise. 
Through this fitting procedure we determine an effective temperature of the cantilever $T=\SI{0.74}{\kelvin}$, which is substantially higher than the bath temperature $T_\mathrm{bath}=\SI{10}{\milli\kelvin}$.
The discrepancy is most likely due to inertial actuation by noise from the piezo shaker and decoupling of the cantilever eigenmode from the thermal bath.

\cref{fig:sensitivity}(b) shows various contributions to the noise of Probe 1 as a function of $\nc$. 
The light blue area marks the range of operation of this sensor, where Kerr nonlinear response of the superconducting resonant circuit~\cite{scarano2024intrinsic} limits $\nc < \num{2.0e6}$, well below $\nc^\mathrm{SQL}\approx \num{1.4e8}$.
It is possible to achieve operation closer to $\nc^\mathrm{SQL}$ by improving the sensor design.

For a given $g_0$ and $\Gamma$ we can decrease $\nc^\mathrm{SQL}$ by reducing the external loss rate $\kex$ and the total loss rate $\kappa$.
We can control \kex through design of the resonator coupling to the readout transmission line~\cite{roos2024design}.  
Decreasing $\kappa$ also requires improvement on the internal loss rate of the cavity $\kappa_\mathrm{0}=\kappa-\kex$. 
With typical parameters for our sensors $\kex/2\pi=\SI{1}{\mega\hertz}$, $\Gamma/2\pi=\SI{10}{\hertz}$, $g_0/2\pi=\SI{0.14}{\hertz}$, we find a value of $n_c^\mathrm{SQL}\approx \num{31.8e6}$ photons, which would require a factor 15 improvement on $\kappa_\mathrm{0}$.
Achieving this improvement is quite reasonable as our microwave resonators exhibit a $\kappa_\mathrm{0}$ far larger than the typical values reported in the literature~\cite{samkharadze2016high,kroll2019magnetic, muller2022magnetic}. 
The main limitation to the internal quality factor of our microwave resonators is the highly-disordered low-stress Si-N layer, which causes significant dielectric losses~\cite{mittal2024annealing}. 

As discussed above, the figure of merit of a displacement detector should be evaluated with respect to the fluctuations of the specific mechanical resonator that is measuring force. 
In this regard, it is informative to define the crossover temperature $T^*$, or effective mode temperature of the mechanical resonator, at which $\Szz^\mathrm{cant}(\Wm)=\Szz^\mathrm{det}$.
When the mechanical mode temperature $T>T^*$, the noise is dominated by mechanical displacement noise and the force measurement is at the thermal limit.
When $T<T^*$, the measurement is dominated by detector noise and improvements to displacement detection will improve force sensitivity.

Piezoelectric sensors with transimpedance amplifier detectors as reported by Giessibl et. al.~\cite{giessibl2011comparison} exhibit $\Szz^\mathrm{det}=\SI{62}{\femto\meter\per\sqrt{\hertz}}$ giving $T^*=\SI{10}{\kelvin}$ for tuning forks, and $\Szz^\mathrm{det}=\SI{1.89}{\femto\meter\per\sqrt{\hertz}}$, giving $T^*=\SI{21}{\kelvin}$ for length-extensional resonators.
Optimized optical-beam deflection detectors as reported by Fukuma \emph{et al.}~\cite{fukuma2005development} exhibit $\Szz^\mathrm{det}=\SI{17}{\femto\meter\per\sqrt{\hertz}}$, $T^*=\SI{1.2}{\kelvin}$. For comparison, from measurements on Probe 1 shown in \cref{fig:sensitivity}(a), we obtain a lower crossover temperature $T^*=\SI{0.43}{\kelvin}$, despite the fact that the sensor operates below its optimal point.

\section{AFM operation}

The noise-to-noise ratio $\Szz^\mathrm{cant}/\Szz^\mathrm{det}=T/T^*$ provides a qualitative measure of the pixel acquisition rate, or measurement bandwidth $\Delta f$, above which force sensitivity is degraded by detector noise.
For FM-AFM this bandwidth is $\Delta f=\Gamma (T/T^*)$ where as AM-AFM requires $\Delta f\leq \Gamma$.
Force sensitivity is however only one aspect of a high-performance AFM.
Ultimately the image speed and resolution, both vertical and lateral, are influenced by many factors: equilibrium tip-surface distance, oscillation amplitude of the driven cantilever, sharpness of the tip, and pickup of environmental vibrations and noise, such as noise from the actuator driving the cantilever oscillation as well as feedback noise and sample topography. 

\begin{figure}
\centering
\caption{Magnitude (a) and phase (b) of the microwave resonator's reflection coefficient $S_\mathrm{11}$ measured during coarse approach. Proximity to the surface causes a non-negligible shift in the microwave resonance frequency (c) and increase in the total loss rate (d). (e) Effect of force gradient on the mechanical susceptibility of the cantilever close to the surface. (f) Mechanical resonance frequency shift vs. tip-surface distance curve.}
\label{fig:approach}
\dblcolfigure{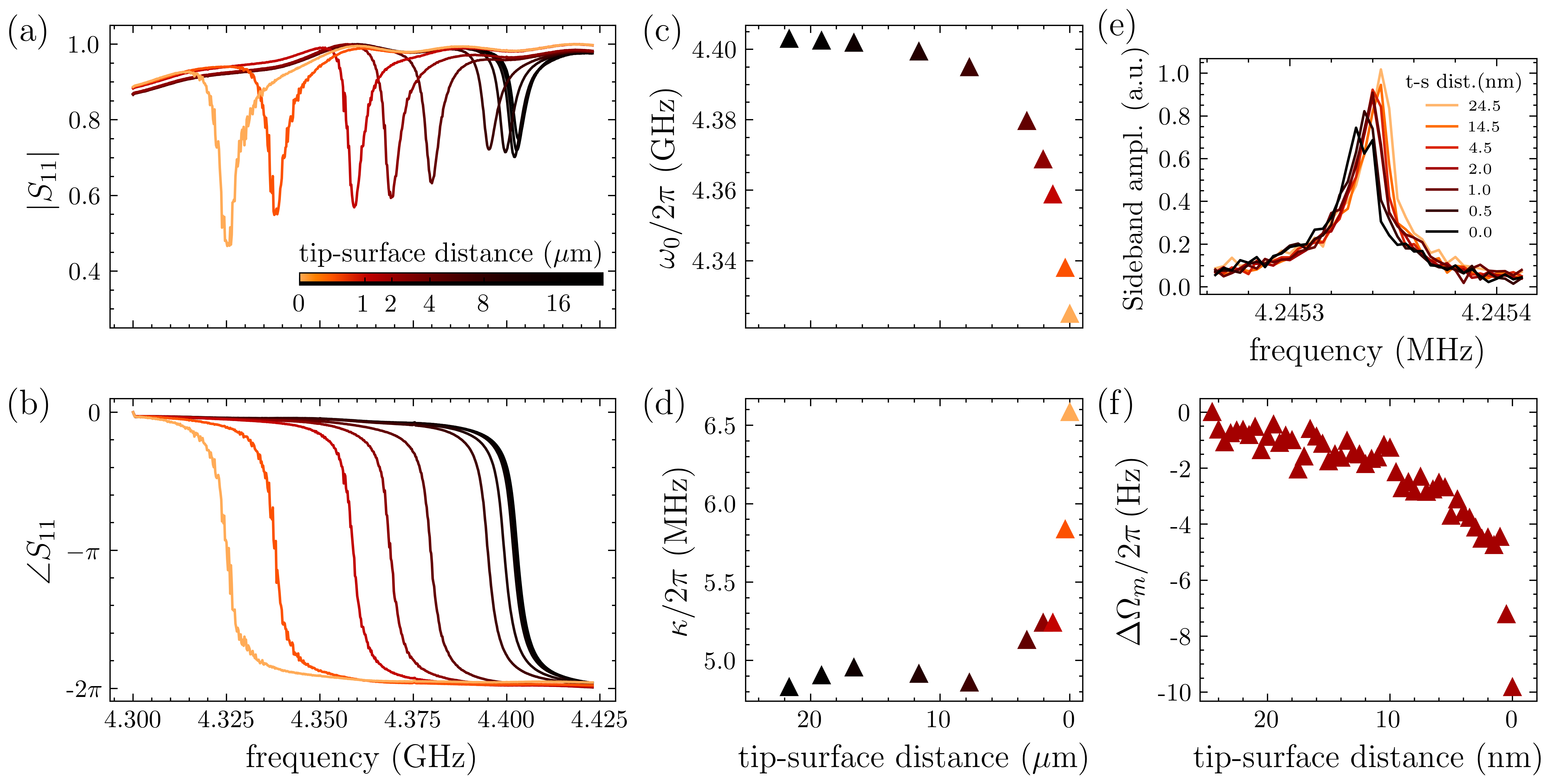}
\end{figure}

We perform initial positioning of the microscope by driving the coarse $z$-positioner with the Attocube ANC-300 control unit that provides the \SI{60}{\volt} saw-tooth pulses to the piezo actuators.
The stick-slip motion of the actuator under high-voltage saw-tooth pulses ($\approx\SI{50}{\nano\meter\per step}$) produces significant heating at the \SI{10}{\milli\kelvin} stage. 
Under continuous operation, the heating power is of the order of \SI{1}{\micro\watt} for an approach speed of \SI[per-mode=symbol]{1}{\micro\meter\per\minute}. 
This temperature change during coarse approach causes drift in the cantilever's resonance frequency, requiring a couple of minutes to restore a stable configuration.
Therefore, during coarse approach we monitor only the microwave resonance.

When approaching the sample surface we observe a shift ($<\SI{2}{\percent}$) of the microwave resonance frequency to lower frequency and a slight increase of the internal losses, as shown in \cref{fig:approach}(a)--(d). 
This red-shift suggests that proximity to the surface introduces a lossy capacitive load to the microwave resonant circuit. 
While slightly degrading the performance of the displacement detector, this effect has the advantage of providing an easily measurable indication of probe-surface separation during coarse approach. 
For the final stage of the approach, we drive the $z$-scanner with a DC-voltage (\SI[per-mode=symbol]{100}{\nano\meter\per\volt}) while monitoring both the microwave and cantilever resonances. \cref{fig:approach}(e)--(f) show the measured frequency shift versus distance, demonstrating force gradient sensing by the KIMEC AFM probe.  

\begin{figure}
\centering
\caption{(a) AM-AFM imaging of the calibration grating over a $\SI{1}{\micro\meter}~\times~\SI{1}{\micro\meter}$ scan area containing a single vertical stripe with a \SI{10}{\nano\meter} step height. The dashed lines in the image mark the location of the scan lines in (b)--(c).}
\label{fig:AM-AFM}
\dblcolfigure{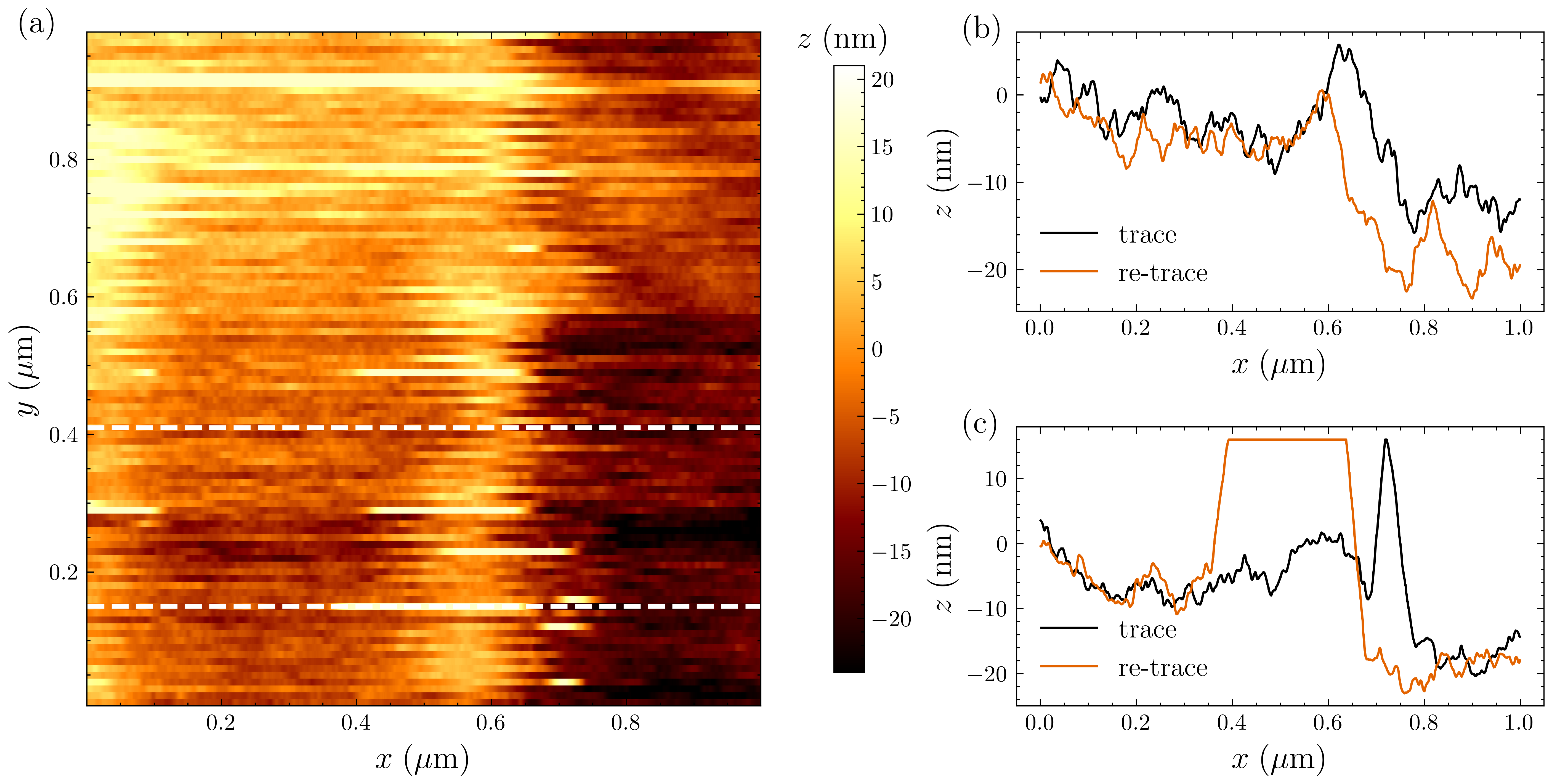}
\end{figure}

We tested the AFM by imaging a nano-fabricated calibration grating over a $\SI{1}{\micro\meter}\times \SI{1}{\micro\meter}$ scan area with a resolution of \SI{100}{pixels\per line}.
In order to track the surface topography while scanning, we developed a proportional-integral-derivative (PID) control firmware running on the embedded real-time processing unit (RPU) of the microwave multifrequency lock-in amplifier (\emph{Presto} by Intermodulation Products AB). The code is released as open source under the MIT license, and is available online in a repository hosted on GitHub~\cite{rpu}. 
The feedback system must be able to follow rapid changes in topography while avoiding feedback oscillation. 
Depending on the mode of operation, AM-AFM or FM-AFM, the PID setpoint and error signal are determined from the amplitude or phase of the motional sideband which is the up-converted cantilever oscillation.

The PID control is developed in the Rust programming language, and features low-pass filtering of the derivative action, as well as clamping of the integrator state to prevent integral windup and of the output control to prevent damage to the AFM probe.

\begin{figure}
\centering
\caption{(a) FM-AFM imaging (trace) of the calibration grating over a $\SI{1}{\micro\meter}~\times~\SI{0.8}{\micro\meter}$ scan area. (b)--(c) Average and standard deviation over ten consecutive acquisitions of a single scan line at constant $y$-position in both trace (b) and retrace (c) scan direction. (d)--(f) Single scan line measured with different pixel acquisition rates $\Delta f$.}
\label{fig:FM-AFM}
\dblcolfigure{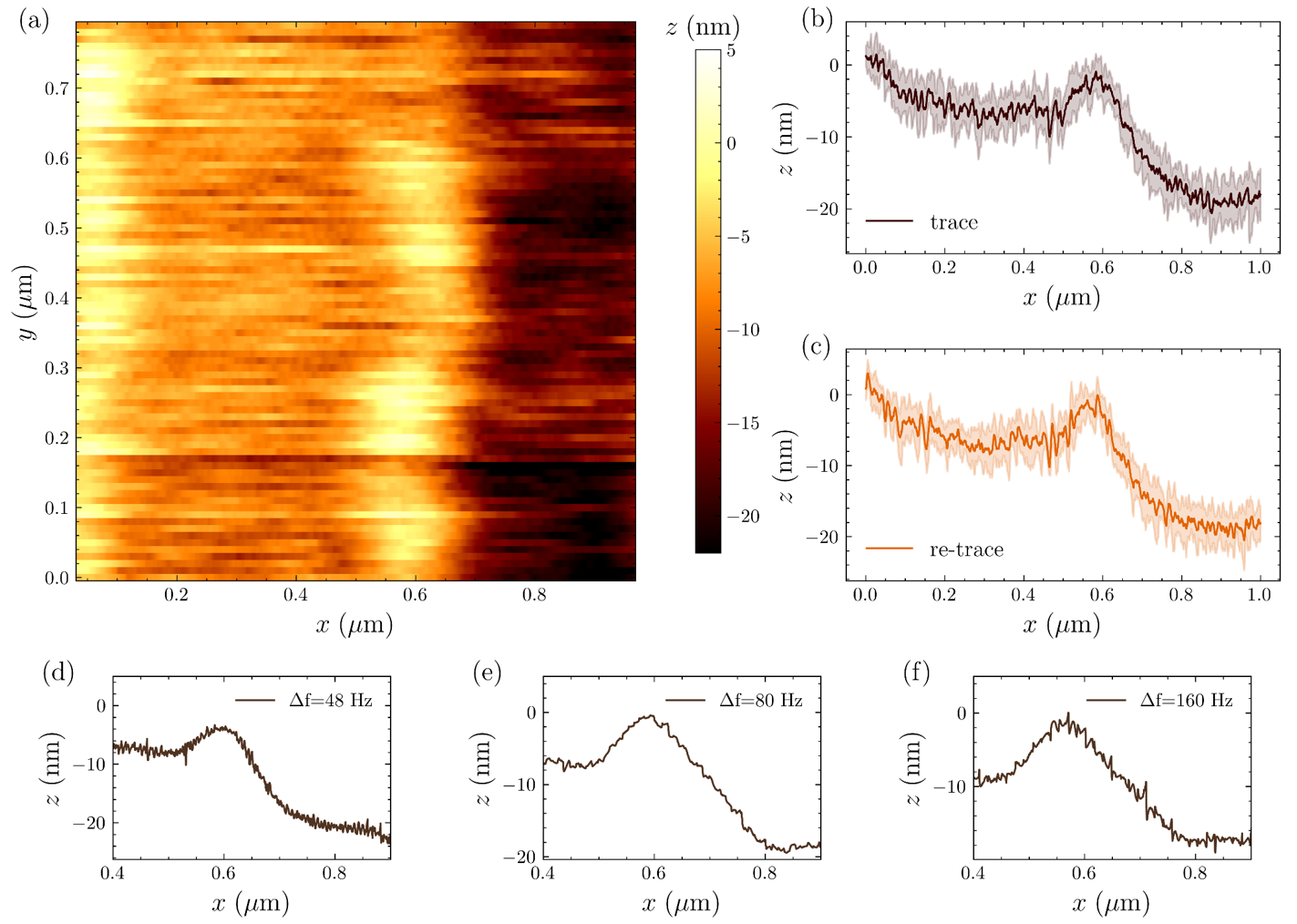}
\end{figure}

We operate Probe 2 in AM-AFM mode by driving the cantilever slightly above resonance $\Omega=\Wm+\Gamma/2$. 
The PID loop controls the $z$-extension of the scanner to keep oscillation amplitude constant at about \SI{200}{\pico\meter}. 
The blue-detuned cantilever drive ensures that the PI controller retracts the $z$-scanner in the event of a sudden drop in the measured signal, thus avoiding crashing the probe into the surface. 
The image in \cref{fig:AM-AFM}(a) is acquired with a measurement bandwidth of $\Delta f=\SI{1.5}{Hz}<\Gamma$, corresponding to a scan rate of roughly \SI{67}{\second \per line}. The choice of the integral gain is crucial for optimal surface tracking. 
Too large values of integral gain result in an overly responsive feedback that introduces noise on the topographic image and feedback oscillations.
Too low values can lead to instability or loss of surface tracking with abrupt changes in topography. 
The latter occurred during the scan; in multiple instances the feedback temporarily loses track of the surface immediately after the tip encounters a step. The single scan line on the fast, i.e. $x$-axis, shown in \cref{fig:AM-AFM}(b)--(c) clearly shows different behavior of trace and retrace scan directions.  

\begin{figure}
\centering
\caption{(a) SEM image of the second test sample. The zoomed inset highlights the area imaged with FM-AFM in (b). (c) Fast Fourier transform of the error signal for a single scan line. The $x$-axis of the plot is converted to frequency through the pixel acquisition rate $\Delta f =\SI{112}{\hertz}$. (d) Single scan line showing the rippling effect caused by pulse-tube vibrations. The orange curve is obtained by applying a low-pass filter with cutoff frequency \SI{0.75}{\hertz}. (e) Effect of filtering and smoothing on the acquired image.}
\label{fig:FM-AFM2}
\includegraphics[width=13.0cm,keepaspectratio]{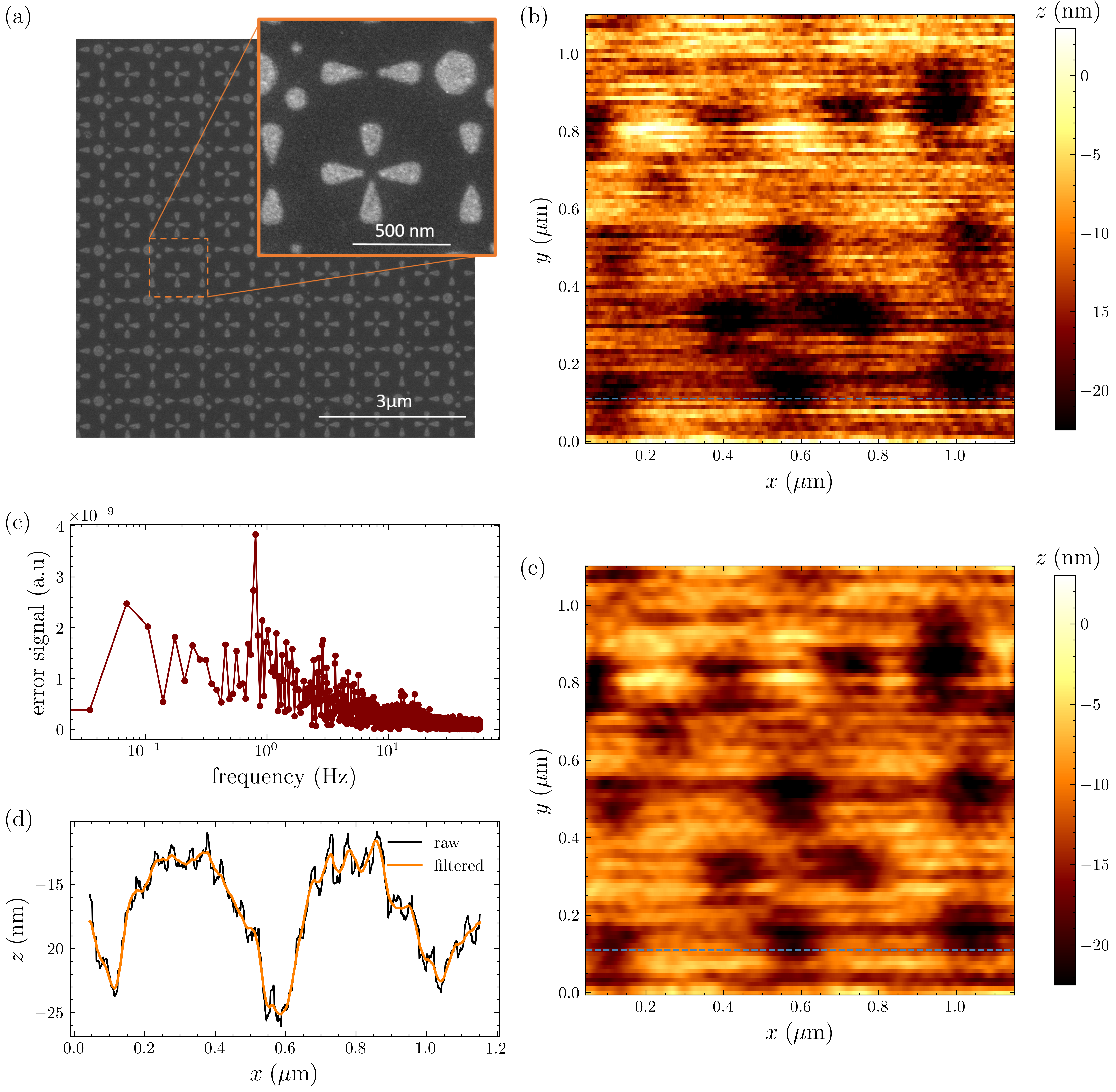}
\end{figure}

We achieved faster imaging and improved stability with FM-AFM as shown in \cref{fig:FM-AFM}. 
FM-AFM requires phase-sensitive detection of tip displacement using a scheme described in detail in earlier publications~\cite{arvidsson2024sensing,roos2023kinetic}.
The cantilever is now driven on resonance at the same $\simeq\SI{200}{\pico\meter}$ oscillation amplitude while the microwave resonator is driven with two pump tones of equal intensity and symmetrically detuned with $\Delta=\pm\Wm$.
The lower and upper motional sidebands from the two pumps interfere at $\omega_\mathrm{0}$, resulting in microwave response amplitude which depends on the phase of the mechanical motion~\cite{Suh2014,Braginsky1980}. 
The feedback controls the $z$-extension to maintain a constant microwave response amplitude, or constant mechanical response phase. 
The tracking of the surface is more consistent using this FM-AFM feedback scheme as shown in \cref{fig:FM-AFM}(b)--(c) by the trace and retrace signals. \cref{fig:FM-AFM}(d)--(f) shows the effect of the measurement bandwidth on a single scan line.

Finally, we tested the lateral resolution of our microscope by imaging a less trivial structure. We fabricated a second test sample by etching a pattern from a \SI{10}{\nano\meter} Ti thin film deposited on a \SI{5}{\nano\meter} Au layer on a silicon substrate. Each unit cell in the pattern contains different shapes as shown in the SEM image in \cref{fig:FM-AFM2}(a). The tapered arms of crosses and ribbons shrink to reach a nominal minimum feature size of \SI{20}{\nano\meter}. \cref{fig:FM-AFM2}(b) shows a FM-AFM image of the $\SI{1}{\micro\meter}\times \SI{1}{\micro\meter}$ scan area corresponding to the inset of \cref{fig:FM-AFM2}(a) acquired using Probe 3. 

Although the various structures can be identified, several scanning artifacts appear in the image. In this particular experiment the probe was especially sensitive to pulse-tube vibrations. \cref{fig:FM-AFM2}(c) shows the fast-Fourier transform (FFT) of the error signal for a single horizontal (fast axis) line scan. The component at \SI{0.8}{\hertz} and its higher harmonics coincide with periodicity of the pulse-tube, resulting in ripples in the height image (feedback signal) as shown in \cref{fig:FM-AFM2}(d). 
In this image we also see distortion and misalignment between consecutive scan lines due to hysteresis and drift in the open loop $x-y$ scanners. These problems are more pronounced when attempting to resolve smaller features, and image processing techniques such as filtering and smoothing are not particularly effective in enhancing resolution, as shown by \cref{fig:FM-AFM2}(e).

\section{Conclusion}

We have demonstrated ultra-low-temperature AFM operation with a kinetic-inductive electromechanical (KIMEC) force sensor. In comparison with state-of-the-art low-temperature AFM, these initial AFM images are not of particularly high resolution. However, our AFM was built in to a standard pulse-tube dilution refrigerator which was not designed for preparation of pristine surfaces, nor was any attempt made to sharpen the tip in-situ.

Our analysis of force sensitivity showed that residual noise from the actuator and decoupling of the cantilever eigenmode from the thermal bath limited the sensor's effective temperature to \SI{0.74}{\kelvin}, well above the bath temperature of \SI{10}{\milli\kelvin}.
However, our analysis shows that the KIMEC principle of detecting cantilever deflection, as it is implemented in these sensors, is capable of operating at the thermal limit of force sensitivity down to a temperature of \SI{0.43}{\kelvin}. 
This is a significant improvement over much lower frequency piezoelectric transduction methods commonly used for low-temperature AFM, which can work at the thermal limit down to some \SI{10}{\kelvin}.  
We estimate that, with reasonable improvements on the current KIMEC sensor design, this detection principle will be capable of operating at the fundamental limit of minimum added noise. 
Achieving this performance limit is a prerequisite for improved force sensitivity with ultra-low temperature and ultra-low damping mechanical modes.  
Alternatively, the lower noise of the KIMEC detection principle allows for AFM imaging with increased pixel acquisition rate (measurement bandwidth) without degrading force sensitivity.

\begin{acknowledgements}
We acknowledge helpful discussions with Thilo Glatzel.
\end{acknowledgements}

\begin{funding}
The European Union Horizon 2020 Future and Emerging Technologies (FET) Grant Agreement No. 828966 --- QAFM and the Swedish SSF Grant No. ITM17--0343 supported this work.
\end{funding}

\section*{Data availability}
Data generated and analyzed during this study is openly available in Zenodo at \url{10.5281/zenodo.15744435}.


\end{document}